\documentclass[a4,12pt]{article}
\usepackage{bm}
\usepackage{amsmath}
\usepackage{epsfig}
\begin{document}
\begin{center}
{\bf \large Scalar response of the nucleon, 
chiral symmetry 
and nuclear matter properties.}\\[2ex]

G. Chanfray$^1$ and Magda Ericson$^{1,2}$, \\
$^1$ 
  IPN Lyon, IN2P3/CNRS, Universit\'e Lyon1, France\\
$^2$ Theory Division, CERN, Geneva, Switzerland

\begin{abstract}
In this talk we present a description of nuclear binding in a chiral approach based on the existence of a chiral invariant scalar field associated with the generation of the masses through spontaneous chiral symmetry breaking. We discuss the emergence of such a field on the example of the NJL model.  We also incorporate the effect of confinement at the level of the nucleon substructure to stabilize nuclear matter. In a particular  quark-diquark model we illustrate the simutaneous influences of  spontaneous chiral symmetry breaking and confinement on the nucleon mass and on the nuclear matter description. 
\end{abstract}
Pacs: 24.85.+p 11.30.Rd  12.40.Yx 13.75.Cs 21.30.-x
\end{center}
\noindent
{\bf Introduction. }
The relation between the fundamental properties of low energy QCD, namely chiral symmetry and confinement, and the nuclear many-body problem is certainly one of the most challenging aspect  of present day nuclear physics. One particular crucial question is how the interplay between 
chiral symmetry and confinement in the nucleon structure manifests itself  in  the nuclear many-body problem. In this talk we will develop  a viewpoint where the nuclear attraction is associated with the in-medium fluctuation of a (chiral invariant) background scalar field which is itself at the origin of the constituent quark mass. Nuclear stability is ensured with the incorporation  of the nucleon  response to this  scalar field. This  response  depends on the quark confinement mechanism inside the nucleon. As we will see this question can be  rephrased in terms of the relative weight of spontaneous chiral symmetry breaking and confinement in the origin of the nucleon mass.

\bigskip\noindent
{\bf Chiral nuclear matter description including the effect of confinement.}
As a starting point let us consider  the relativistic mean-field approaches initiated by Walecka where the nucleons move in an attractive scalar  and a repulsive vector background fields. It provides an economical saturation mechanism and a well known success is the correct magnitude of the spin-orbit potential where  the large vector and scalar fields contribute  in  an additive way. Now the question of the very nature of these background fields  and their  relationship with the QCD condensates has to be elucidated. To address this question we formulate an effective theory     parametrized in terms of the fields associated with the fluctuations of the chiral condensate in a matrix form (${\cal M}=\sigma + i\vec\tau\cdot\vec\pi$) 
by going from cartesian  to polar coordinates,  {\it i.e.}, going from a linear to a non linear representation~:
${\cal M}=\sigma\, + \,i\vec\tau\cdot\vec\pi=S\,U=(f_\pi\,+\,s)\,exp\left({i\vec\tau\cdot\vec\Phi_\pi/ f_\pi}\right)$.  In ref. \cite{CEG02} we made the physical assumption  to identify the chiral invariant scalar field, $s=S - f_\pi$, associated with {\it radial} (in order to respect chiral constraints) fluctuations   of the condensate, with the background attractive scalar field. In this picture the nuclear medium can be seen as a shifted vacuum characterized by a chiral order parameter $S= f_\pi + s$ which governs the evolution of the nucleon mass ($\Delta M_N/M_N=\left\langle s\right\rangle/f_\pi)$ and part of the evolution of the chiral condensate. It  turns out that this scalar field, which  has derivatives coupling with the pion, largely decouples from low energy pions whose dynamics is governed by chiral perturbation theory. This scalar field is relevant for low momentum nuclear physics and the question of its relation with  the famous $\sigma(600)$ (which appears experimentally as a broad $\pi\pi$ resonance) is to a large extent irrelevant for our purpose.

In our approach the Hartree energy density of nuclear matter (including omega exchange) writes in terms of the order parameter $\bar s=\langle s \rangle $~: 
$\quad E_0/ V=\varepsilon_0=\int\,(4\,d^3 p /(2\pi)^3)$\break$\Theta(p_F - p)\,E^*_p(\bar s)\,+\,W(\bar s)\,+\,g^2_\omega/2m_\omega^2\,\rho^2,$
where $E^*_p(\bar s)=\sqrt{p^2\,+\,M^{*2}_N(\bar s)}$ is the energy of a nucleon with the effective Dirac mass, $M^*_N(\bar s)=M_N +g_S\,\bar s$ and  $g_S=M_N/f_\pi$ is the scalar coupling constant of the sigma model. Here two serious problems appear. The first one is  the fact that the chiral effective potential, $W(s)$, contains an attractive tadpole diagram which  generates an attractive three-body force destroying  matter stability \cite{BT01}. The second one is related to the nucleon substructure. According to the lattice data analysis of  Thomas et al \cite{LTY04}, the nucleon mass can be expanded according to $M_N(m^{2}_{\pi}) = a_{0}\,+\,a_{2}\,m^{2}_{\pi}\, +\,a_{4}\,m^{4}_{\pi}\,+\,\Sigma_{\pi}(m_{\pi}, \Lambda)+...$, where the pionic self-energy is explicitely separated out. While the $a_2$ parameter is related to the non pionic 
piece of the $\pi N$ sigma term,  $a_4$ is related to the nucleon QCD scalar susceptibility. The important point is that   $(a_4)_{latt} \simeq- 0.5\, \mathrm{GeV}^{-3}$ \cite{LTY04} is essentially compatible with zero in the sense that it is  much smaller than in our chiral effective model, $\left(a_4\right)_{Chiral}=-3f_{\pi} g_S/2\,m^{4}_{\sigma}\simeq -3.5\,GeV^{-3}$ , where the nucleon  is seen  as a juxtaposition of three constituent quarks getting their mass  from the chiral condensate \cite{CE05}. The common origin of these two failures can be attributed to the absence  of confinement \cite{MC08}. In reality the composite nucleon  responds to the nuclear environment, {\it i.e.}, by readjusting its confined quark structure \cite{G88}. The resulting  polarization of the nucleon is accounted for by the phenomenological introduction of the {\it positive} scalar nucleon response, $\kappa_{NS}$, in  the nucleon mass evolution ~: $M_N(s)=M_N\,+\,g_S\,s\,+\,\frac{1}{2}\,\kappa_{NS}\,s^2\,+\,....$.
This constitutes the only change in the expression of the energy density  but this has numerous consequences. In particular
the $a_4$ parameter is modified~: $a_4=\left(a_4\right)_{Chiral}\left(1\,-\,\frac{2}{3} C\right)$. The value of  $C\equiv(f^{2}_{\pi}/2\,M_N)\kappa_{NS}$ which reproduces the lattice data is $C\simeq 1.25$ implying a strong cancellation effect in $a_4$
\cite{CE05, CE07}. Moreover  the scalar response of the nucleon induces an new piece in the lagrangian ${\cal L}_{s^2 NN}=-\,\kappa_{NS}\,s^2\,\bar N N/2$ which  generates a repulsive three-body force able to restore saturation \cite{EC07}.

The restoration of saturation properties has been confirmed at the Hartree level \cite{CE05} with a value of the dimensionless scalar response parameter, $C$, close to the value estimated from the lattice data, taking   $g_S=M_N/f_\pi$ and $g_\omega$ adjusted near  the VDM value. The next step has been to include pion loops on top of the Hartree mean-field calculation \cite{CE07,MC09}. One possibility is to use in-medium chiral perturbation theory but we prefer to use a standard many-body (RPA) approach which includes the effect of short-range correlations ($g'$ parameters fixed by spin-isospin phenomenology) and  $\Delta-h$ excitations.   The calculation which also incorporates rho exchange has no real free parameters apart for a fine tuning of $C$ (around the lattice estimate) and of  $g_\omega$  (around the VDM value). We have The inclusion of spin-isospin  loops improves the quality of the result,  in particular  they bring down the compressibility close to the accepted value. We stress   the relatively modest value of the correlation energy ($\simeq -10\,MeV$), much smaller than what is obtained from iterated pion exchange (planar diagramm) in  in-medium chiral perturbation theory. This effect is mainly due to the strong screening of pion exchange by short-range correlations.

We have also performed  a full relativistic Hartree-Fock calculation with the notable inclusion of the rho meson exchange which is important to also reproduce the asymetry properties of nuclear matter \cite{MC08}. Again in an almost parameter free calculation, saturation properties of nuclear matter can be reproduced with $g_S=10$, $m_\sigma=800\, MeV$ , $g_\rho=2.6$ (VDM), $g_\omega=6.4$ (close to the VDM value $3\,g_\rho$) and $C=1.33$ (close to the lattice value).   An  important conclusion is that  the rho Hartree contribution ($7\,MeV$) to the asymmetry energy $a_S$ is not sufficient to reproduce the accepted value of $a_S$ around $30\, MeV$   when keeping the VDM value for the vector coupling constant, $g_\rho$. The Fock term through its tensor contribution is absolutely necessary and the agreement is even better when   $\kappa_\rho$ is increased from the VDM value, $\kappa_\rho=3.7 $ to $\kappa_\rho=5 $ (strong rho scenario). The model also predicts a neutron mass larger than the proton mass with a difference increasing  with neutron richness in agreement with ab-initio BHF calculations \cite{SBK05}.

\bigskip\noindent
{\bf Status of the nuclear scalar field.}
Since the very notion of a scalar field has met some septicism it is important to remind how such an object naturally emerges. For this purpose we introduce the NJL model in  the light quark sector limited to describe the main mesons~:  the pion, the sigma, the rho, the $a_1$ and the omega mesons. The  lagrangian is~: \\
$$
{\cal L}= \bar{\psi}\left(i\,\gamma^{\mu}\partial_\mu\,-\,m\right)\,\psi+\,(G_1/2)\,\left[\left(\bar{\psi}\psi\right)^2\,+\
\left(\bar{\psi}\,i\gamma_5\vec\tau\,\psi\right)^2\right]
$$\break$$-\,(G_2/2)\,\left[\left(\bar{\psi}\,\gamma^\mu\vec\tau\,\psi\right)^2\,+\,
\left(\bar{\psi}\,\gamma^\mu\gamma_5\vec\tau\,\psi\right)^2\,+\,\left(\bar{\psi}\,\gamma^\mu\,\psi\right)^2\right].
$$
\\
It is generally accepted that this model gives an excellent description of vacuum chiral symmetry breaking, with the appearence of a constituent quark mass and a soft Goldstone pion mode. In addition in the so called delocalized version the inclusion of  a form factor, depending on a cutoff, $\Lambda$, at each quark leg of the interaction generates a momentum dependence of the quark mass in agreement with lattice calculation. However the approach  is notoriously not fully satisfactory due to the lack of confinement~: in particular unphysical decay channels of vector mesons in $q \bar q$ pairs may appear. In that respect  it has been shown that adding a confining interaction on top of the NJL model solves the problem of unphysical $\bar q q$ decay channels of mesons with masses larger than twice the constituent quark mass \cite{CSWSX95}. In addition the scalar meson arising around twice the constituent quark mass is pushed at  higher energy by confinement and the $\sigma(600)$ comes out as a broad $\pi\pi$ resonance. It has also been demonstrated that the effect of the confining interaction, although crucial for the on-shell properties of vector and scalar mesons, has only little influence for the low momentum physics  relevant for the nuclear many-body problem \cite{CWS01}. Hence, our attitude will be to derive an effective low momentum theory where the mass parameters for scalar et vector mesons will not be the on-shell mesons masses but simply mass parameters associated with the inverse of the corresponding correlators taken at zero momentum. Technically this can be done by  rewriting the NJL lagrangian in a semi-bozonized form
and integrating out quarks in the Dirac sea using a path integral formalism. The physical meaning is simply a projection of $q \bar q$ vacuum fluctuations onto meson degrees of freedom. Keeping only the relevant terms for nuclear physics purpose, the resulting low momentum effective lagrangian has the form~:
\begin{eqnarray}
{\cal L}&=&\frac{1}{2}\,\frac{I^l_{2S}(\bar{\cal S})}{I^l_{2S}(M_0)}\partial^{\mu} S\partial{\mu}S\,-\,W	({\cal S}=g_{0S} S)
+\frac{1}{4}\,F^2\,M_\pi^2\,\frac{\bar{\cal S}}{M_0}\,tr_f(U\, +\, U^\dagger\,-\,2)\,\nonumber\\
& &+\,
\frac{1}{2 F^2}\tilde I(\bar{\cal S})\,\bar{\cal S}^2\,\partial^{\mu}\vec{\Phi}\partial^{\mu}\vec{\Phi}
+ \frac{1}{2}M^{2}_{V}\,\left(\omega^{\mu}\omega_{\mu}\,+\,\vec{v}^{\mu}\cdot\vec{v}_{\mu}\right)\,+\,
\frac{1}{2}M^{2}_{A}\,\left(\vec{a}^{\mu}\cdot\vec{a}_{\mu}\right)
\nonumber\\
& &-\,\frac{1}{4}\,\left(\omega^{\mu\nu}\omega_{\mu\nu}\,+\,\vec{v}^{\mu\nu}\cdot \vec{v}_{\mu\nu}\,+\,\vec{a}^{\mu\nu}\cdot \vec{a}_{\mu\nu}\right)
\end{eqnarray}
${\cal S}$ is a chiral invariant scalar field whose vacuum expectation value coincides with the constituent quark mass $M_0$ and $S$ is the canonical one. We also introduce an effective scalar field  $(S)_{eff}=(F_\pi/M_0){\cal S}$  normalized to $F_\pi$ in the vacuum as in our previous approaches and  $W$ is the corresponding chiral effective potential. The matrix $U$ is $U=exp(\Phi/F)$ where $\Phi$ is the orthoradial pion field and the parameter $F$ can be identified with the pion decay constant parameter $F_\pi$. The quantity
$\bar{\cal S}=(M_0/F_\pi)(F_\pi +\bar s)$ is the expectation value of the scalar field which is expected to decrease in the nuclear medium. The various $I$'s functions are standard NJL loop integrals which also enter the mass parameters $M_V$ and $M_A$  for vector and axial-vector mesons. 
These masses also depend on the vector coupling constant $G_2$. In particular the ratio $g^{2}_{V}/M^{2}_{V}$ ($g_V$ being the quark-vector coupling constant) is fixed to $G_2$. Finally the pion mass emerges as $M^{2}_{\pi}=m\,M_0/G_1\,F^2_\pi$.
We have a priori four parameters,  $G_1, G_2$, the cutoff  $\Lambda$ and the bare quark mass $m$.  We use~: $ \Lambda=1\, GeV, \quad m=3.5\, MeV, \quad G_1=7.8\, GeV^{-2}.$
We thus obtain for the vacuum quark mass at zero momentum~:
$M_0= 371\, MeV$ and $\left\langle \bar q q\right\rangle= -(286\, MeV)^3$. We actually constrain $G_2$ to be close to the VDM value
$(G_2)^{VDM}= g^{2}_{V}/M^{2}_{V}=(2.65/0.770)^2\,GeV^{-2}$. For the nuclear matter calculation we take $G_2=0.78\,(G_2)^{VDM} \Rightarrow F_\pi=93.6 MeV, \quad M_\pi=137.8\, MeV  $. With this set of parameters the low momentum mass parameters are~:
$ (M_\sigma)_{eff}=659\, MeV$,  $M_V=653\, MeV$  and $M_A=797\, MeV$.

\noindent
\begin{figure}
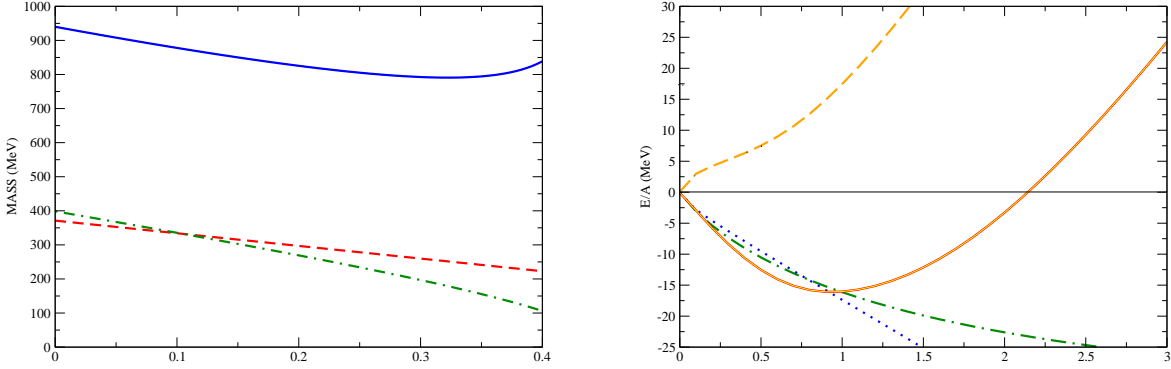

  \begin{tabular}{cc}
  \begin{minipage}{.50\linewidth}
    \includegraphics[scale=0.3]{delocmass.eps}
  \end{minipage}
  &
  \begin{minipage}{.50\linewidth}
\includegraphics[scale=0.3]{delocbind-2.eps}
   \end{minipage}
   \end{tabular}  
\caption{Left panel: Mass of the quark (dashed line), of the diquark (dot-dashed line) and of the nucleon (full line) versus the relative deviation of the scalar field with respect to its vacuum value. Right panel: Binding energy of nuclear matter versus nuclear matter density in units of normal density. 
The full line corresponds to the full result and the dashed line represents the  Hartree result. 
The dot-dashed line corresponds to 
the contribution of the Fock term and the  dotted line  represents the
correlation energy. All the numerical inputs are given in the text.}
\end{figure} 

\bigskip\noindent
{\bf Effect of confinement~: a toy model for the nucleon.}
We now come to the second imortant point of this paper, namely the scalar response of the nucleon. Clearly a nucleon made of the simple adjonction of three NJL constituent quarks will not exhibit a scalar response since the confinement mechanism is lacking. The question of the scalar response is intimately related to the respective role of chiral symmetry breaking and confinement in the generation of the nucleon mass. One possibility is to reduce the relative role of chiral symmetry breaking by considering a nucleon made of a quark and a sufficiently light diquark to leave room for the confinement. As discussed in a set of works of Bentz et al (see ref. \cite{BT01}), it is possible to construct   from the NJL model a nucleon  which  has a diquark component.  The diquark mass   is also medium dependent since it depends on the constituent quark mass.  If   we take for the the coupling constant in the diquark channel, 
$\tilde{G}_1=0.92\, G_1$, we obtain for the vacuum diquark  $M_D= 398.5\, MeV$
which turns out to be nearly equal to the constituent quark mass. In \cite{BT01}, a nucleon scalar response is obtained through the inclusion of  an infrared cutoff $\mu_R\simeq\, 200 MeV$  in the Schwinger proper time regularization scheme. Such a prescription implies that  quarks cannot propagate at relative distance larger than $1/\mu_R$, hence mimicking a confinement mechanism. Here we propose to incorporate confinement in a more direct way. Since the diquark is in an anti-triplet color state, it is physically plausible that a string develops between the quark and the diquark as in a $Q\bar Q$ meson. We thus introduce a confining potential between the quark and the diquark: $V(r) = K\, r^2$.
In the non relativistic limit, the problem reduces to solve the Schrodinger equation for a particle with reduced mass $\mu$, placed in an harmonic potential. In this limit the mass of the (in-medium) nucleon is given by~:
$$M_N(\bar{\cal S})=M(\bar{\cal S})\,+\, M_{D}(\bar{\cal S})\,+\,\frac{3}{2}\sqrt{\frac{K}{\mu(\bar{\cal S})}}\qquad\hbox{with}\qquad
\mu=\frac{M\, M_D}{M + M_D}$$
Taking for the string tension a standard value $K=(290 \, MeV)^3$, we obtain for the vacuum nucleon mass $M_N=1304\, MeV$. Thus the nucleon mass   receives about equal components from   chiral symmetry breaking  and  confinement.
The scalar coupling constant of the nucleon to the effective scalar field (normalized to $F_\pi$ in the vacuum) is
$(g_S)_{eff}(\bar{\cal S})=(M_0/F_\pi)(\partial M_N/\partial \bar{\cal S})_{\bar{\cal S}=M_0}$.
 Its vacuum value is $(g_S)_{eff}=7.14$. What really matters for the magnitude  of the scalar attractive potential  is the ratio $(g_S)_{eff}/(M_\sigma)_{eff}=7.15/659 MeV$ which is close to the value inspired from the linear sigma model used in or previous works,  $10/850 MeV$. The other important point is that $(g_S)_{eff}$ decreases when the scalar field deviates from its vacuum value, {\it i.e.}, when $s=S-F_\pi$ increases in absolute value at finite density. This translates into the fact that the nucleon masses stabilizes and even starts to increase with increasing nuclear scalar field $s$ (see fig. 1).
 The pion nucleon sigma term can be obtained using the Feynman-Hellman theorem. Its numerical value is
$\sigma^{(no pion)}_{\sigma}= 30\, MeV$.
Of course it also receives another contribution from the pion cloud. According to our previous works \cite{CE07} we expect:
$\sigma^{(pion cloud)}\simeq 20\, MeV$ which corresponds to a pion cloud self-energy of $- 420\,MeV$. This is the order of magnitude needed to decrease the nucleon mass from $1.3\, GeV$ to the physical value. In the following we will substract
the nucleon mass by $1304-940=364\, MeV$ which we attribute to the pion cloud. In fig.1 we show the result of the calculation of the various masses versus  $\bar s/F_\pi$. 

We are now in position to calculate the energy of symmetric nuclear matter in the Hartree approximation. The energy density  writes~:
\begin{equation}
{E_0\over V}=\varepsilon_0=\int\,{4\,d^3 p\over (2\pi)^3} \,\Theta(p_F - p)\,\left(\sqrt{p^2+M^{2}_{N}(\bar{\cal S})}
\,-\,(M_N)_{vac}\right)
\,+\,W(\bar{\cal S})\,+\,9\,\frac{G_2}{2}\,\rho^2.
\end{equation}
The expectation value for the scalar field is obtained by minimization of the energy density, 
$\partial\varepsilon_0/\partial \bar{\cal S}=0$.
The resulting curve (fig. 1)  displays a saturation mechanism driven by the scalar nucleon response but the binding is not sufficient unless we decrease artificially the vector coupling constant $G_2$ at a value much smaller than the VDM result.  We add on top of the Hartree mean field result the pion loop (Fock term and correlation energy) contribution obtained in our previous work \cite{CE07}. Taking 
$G_2=0.78\,(G_2)^{VDM}$, we obtain a very decent saturation curve shown in fig. 2. Of course a fully consistent calculation 
within the model of the pion loop energy would certainly modify the result but a fine tuning on $G_2$ would presumably be
sufficient to recover the correct saturation curve. The lesson of this simple model calculation seems to confirm our previous conclusions. The confinement effect (scalar response of the nucleon) is able  to stabilize nuclear matter and the pion loop correlation energy helps to get the correct binding energy. We finally stess that the whole approach relies on the existence of a chiral invariant scaler field, not reducible to an effective two-pion state, but intimately related to the generation of the masses through spontaneous chiral symmetry breaking.

\smallskip\noindent
{\it Acknowledgments.}  It is a pleasure to mention that many ideas developped in this talk have been influenced by the works of Tony Thomas on the  lattice analysis of the nucleon  mass and on the implications of the QCD nucleon  structure on the nuclear many-body problem.


\end{document}